# THE GLOBAL OSCILLATIONS OF THE SUN FROM TSI, SOHO AND SDO DATA

[1]Efremov V.I., [2]Parfinenko L.D., [1,2]Solov'ev A.A.

[1]Central (Pulkovo) Astronomical Observatory of the Russian Academy of Sciences (Russia, St. Petersburg)

[2]Kalmyk State University, (Elista, Russia)

Abstract

The global oscillations of the Sun are investigated on the base of three independent sets of data:
1. Records of Total Solar Irradiance (TSI).
2. The average brightness of the photosphere in MDI (SOHO) and HMI (SDO) images.
3. Records of the time variations of sunlight reflected from the planets during their passage in the field of view of the LASCO C3 coronagraph.

It is found that in the low-frequency spectrum of solar oscillations as a star there are modes of 8-10, 36-38 hours and the main stable 9-day mode. The last mode can be regarded as a reflection of oscillatory processes near the fast rotating core of the Sun according to Fossat et al. (2017).

1. Introduction

The paper studies the global oscillations of the Sun in the period range from several hours to several hundred hours. A priori it is clear that such oscillations are to be very small in amplitude. Indeed, until recently there was no convincing evidence of their existence. The detection of global solar oscillations is determined by the progress of observational technologies. Before the appearance of satellite data, a reliable solution to this problem was almost impossible. For the first time about the observations of "solar pulsations" by the ground-based instruments was reported in Severnyi et al. (1976 ), but later these results were not confirmed.

Long-period global oscillations of the Sun can be regarded as an external manifestations of oscillatory processes in the zone of the solar core. Kennedy et al. (1993) suggested that g-modes can serve as modulators of p-mode frequencies, appearing as side lobes in frequency profiles of p-modes. V. Nakaryakov advanced the idea that long-period oscillations of active solar formations, as well as fluctuations in the brightness of the Sun as a star, can be caused by the influence of global gravitational modes (Foullon et al. 2011).

One of the main goals of the heliosysmological missions GOLF, VIRGO and MDI/SOI on board SOHO was the detection of long-term oscillations just as solar gravity modes (Gabriel et al., 1995). Recently Fossat et al. (2017) using a 16-year data series of a special instrument GOLF/SOHO claimed that: "the previously unknown g-mode splitting have now been measured from a non-synodic reference with very high accuracy, and they imply a mean weighted rotation of 1277 +/-10 nHz (9-day period) of their kernels, resulting in a rapid rotation frequency of 1644+/- 23 nHz (period of one week) of the solar core itself, which is a factor 3.8+/- 0.1 faster than the rotation of the radiative envelope". This gradient of the angular velocity of rotation is much greater than in the tachocline at the bottom of the convective zone, so the "winding" of the magnetic field will occur, in the main, not in the tachocline, but much deeper, at the bottom of the radiative envelope. This fact may give rise to a new understanding of the solar dynamo mechanism.

In this paper, we obtained a period close to 9 days in long-period variations of the brightness of the Sun as a star, as well as in TSI oscillations. It is possible that these oscillations are associated with the g-modes generated in the solar core. If this is the case, then there is an additional opportunity to investigate the echoes of processes in the Sun's core by the available means.

## 2. Investigation of long-period oscillations of the Sun as a star from measurements of Total Solar Irradiance (TSI)

Total solar radiation (TSI) is a measure of solar energy over all wavelengths per unit area falling on the upper atmosphere of the Earth. It is measured on a flat surface oriented perpendicular to the incoming sunlight. The "solar constant" is a common measure of the average TSI at a distance of one astronomical unit.

The energy of TSI largely determines the climate of the Earth. Satellite observations showed that the total solar radiation (TSI) varies slightly during the solar cycle, within 0.1%. At short intervals, the change is greater. We are not interested in random changes in TSI caused by sunspots, faculae, etc., but possible periodic TSI pulsations in the period range from several hours to several hundred hours. They can be due to the fact that the process of transferring of radiation from the core of the Sun to the photosphere, established over billions of years, can undergo a certain modulation with a main period of about 9 days in accordance with the results of Fossat et al. (2017).

Here, we use the data for TSI obtained by the SOlar Radiation and Climate Experiment (SORCE) - http://lasp.colorado.edu/home/sorce/data/tsi-data/ . SORCE carries four instruments, including the spectral irradiance monitor (SIM), solar stellar irradiance comparison experiment (SOLSTICE), total irradiance monitor (TIM), and the XUV photometer system (XPS). In our study, we used the daily data of TIM. TIM measures the total solar irradiance (TSI) with an estimated absolute accuracy of 350 ppm (0.035%). Relative variations in the solar irradiance were measured with the accuracy less than 0.001%/yr, Kopp (2011).

**2.1. Methods of the treatment and observational data**

To determine the spectral composition of the selected time series, we use CaterPillarSSA technique (Golyandina et al. 2001 ), developed recently for the study of nonlinear systems and non-stationary time series. In the literature, the method is best known as SSA (Singular Spectrum Analysis) (Broomhead, King, 1986). The method is successfully used in the physics of the Sun, geophysics, medicine, astrophysics and other branches of science. Without going into details of the method, we note some of its positive features. So, unlike the traditional methods of spectral analysis, this method uses directly the investigated time series to form a decomposition basis, with its subsequent transformation into a trajectory matrix and the orthogonalization of the matrix using the procedure of SVD (Singular Value Decomposition). The undoubted advantages of the method are that it is possible to obtain the "instantaneous" frequencies of the time series, i.e. there is the ability to locally and adaptively describe the frequency components for almost any oscillating signal. In the program implementation of this method, the main components of the decomposition can be evaluated and, which is especially important in the further reconstruction of the original series, they can be visualized and ordered in increasing their contribution to the initial series. The phase diagrams directly show the nature of the component's circulation process and make it possible to assemble the series, excluding one or the other component. This allows us to interactively produce a direct search for harmonic components, filtering or smoothing a series. An important part of the research process is the construction of a correlation matrix of decomposition components to analyze their independence. The matrix allows us to combine the decomposition components into an independent oscillatory mode of the process. Since the real oscillatory mode of the signal can be far from the harmonic form, the requirement of the independence of the modes (and not the components!) constituting the original time series is a natural physical limitation imposed on the nonlinear process under study. The main parameter of the basic algorithm of SSA is the length of the window "L" (from "Lag"). The problem is the effect of the window length on the "separability" of the decomposition components, i.e. on the orthogonality of the corresponding time series (the independence of the modes of the expansion) and on the proximity of the eigenvalues. There are no general universal rules for selecting the length of the "L" window, but there are some recommendations. So, for example, to reliably identify the trend component, the "L" must be large enough, up to half of the series (~ N / 2), and to extract the periodic components - a multiple of them. Here there is some uncertainty: on the one hand, a large window length improves the "separability" of components, but greatly increases the computational resource, and on the other hand a small window length can lead to mixing of the interpreted components of the series, which, of course, should be excluded. Thus, the variation of the parameter "L" is a subtle tool in the study of time-varying time series, where the intuition and practical experience of the researcher is often the decisive factor. This makes the CaterPillarSSA method more attractive, for example, before Hilbert-Huang Transform (HHT), in which the number of components is determined by the automatic output of their iterative process.

By this way, the aim of the method is to expand the time series into interpretable additive components with further visualization of its component modes. The method does not require the stationarity of the series, the knowledge of the trend model, as well as information on the availability of periodic components and their periods. As a complement to SSA, we will also perform a wavelet transformation for the time series of TSI (Astaf'eva, 1996). In recent years, this is a rather traditional method of spectral analysis, using a pre-selected analytical basis for the decomposition. We will construct a wavelet map using a continuous wavelet transformation, where as the pattern we take the Morle wavelet of the 5th order (CWT, Morlet5th). Often, this approach is justified when considering the overall dynamics of components (component drift across the spectrum, merging and separation).

As an observational material for detecting of the 9-day mode in the TSI variation series, we take several randomly selected non-crossing fragments from the total record of TSI (SORCE Total Solar Irradiance - Six Hour Average- Time Series; http://lasp.colorado.edu/lisird/data/sorce_tsi_6hr_l3/).

For our problem, the series with a duration on the order of 5-10 of the required periods are suitable: ~ 50-100 days (2-3 months). We analyze in more detail one observation, showing all the steps of the SSA method, while for others we give only the results of the study (see Table 1).

### 2.2. Observation of TSI

The Julian date are: from 2457849.625 to 2457929.375; i.e. 2017, from 5 April to 25 June; the observation duration is $T = 80$ days, $N = 320$ pts (points of time series), the cadence of the series is equal to 6 hours, and the period of the expected mode $P = 9$ days ($n = 36$ pts). A general view of the time variation of the TSI flux is shown in Fig. 1.

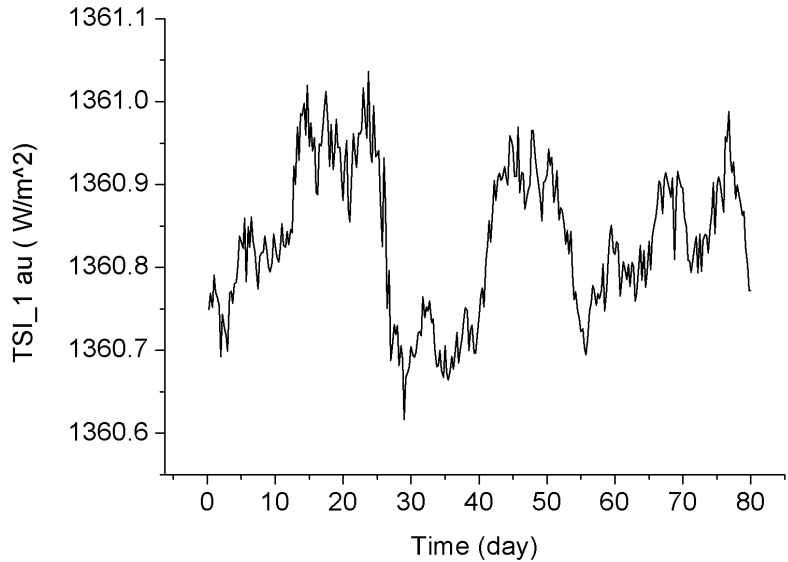

Fig.1. Change in the TSI flux during the observation time.

We will consider our time series as a sum of three additive components: trend, periodicals and noise. Without going into the mathematical details of the method, we will use its software implementation (Golyandina et al. 2001 ) and perform SVD decomposition of the original time series. Let L = 144. In accordance with the above recommendations (on the one hand, L ~ N / 2, that is, it is large enough for a confident trend isolation, and on the other hand L = k * P is a multiple of the expected mode of the series, where k is the whole number). The decomposition shows the presence of a small trend in this part of the TSI series and the "strong" (large amplitude) periodic component of the period of 27.5 days; therefore, for further investigation of the periodicals of smaller amplitudes, we filter out both the trend and this component. Fig. 2a shows the sequence of singular numbers of the trajectory matrix (a matrix constructed from successive fragments of the time series under study) arranged in descending amplitude values. The periodic component of the series has a complex character: it consists essentially of three components of the decomposition (the three "pairs" in the left part of Fig. 2a), and the singular numbers with n ≥ 7 represent the noise block (monotonic decrease of the values of EVAL (eigen values), this set is generated by the noise component of the series.

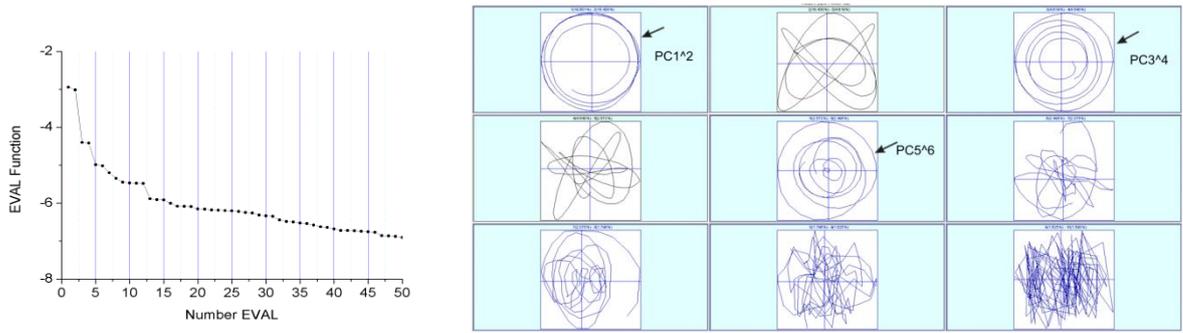

Fig. 2. Left (2a): the logarithms of the first 50 eigenvalues; On the right (2b): the first 9 phase diagrams for the corresponding pairs of eigenvalues, also arranged in descending order of amplitudes. The arrow indicates the Principal Components (PC) that show the circulation nature of the process with periods of 9 days (the pair PC with eigen values 1, 2), 6.6 days (PC with 3,4), and 4.7days (PC with 5,6), respectively.

On the right hand site of Fig. 2b the phase diagrams of the main components (PC-principal components) for the corresponding pairs of eigenvalues of the expansion, also in order of decreasing amplitude as in Fig. 2a, are presented. In fact, these are two-dimensional images of eigenvectors, the diagrams of which confirm the circulatory nature of the decomposition components (ideally, a circle for the harmonic component). Note that for the correct construction of the oscillatory component of the time series (the sought mode can consist of several components of the decomposition in a complex manner), the structure of singular vectors is more significant than the knowing the magnitudes of the eigenvalues. An essential factor is also the independence of the components that make up the group, which is determined by the correlation matrix. Let us consider the correlation relations of the components of the expansion. Fig.3 shows the matrix of correlations of the recovered components. It can be seen that the groups of components along the diagonal of the matrix, which are collected in accordance with the distribution of the eigenvalues of the expansion (Fig. 2a), form independent (orthogonal) modes. As expected, the components with $n \geq 7$ form a noise group.

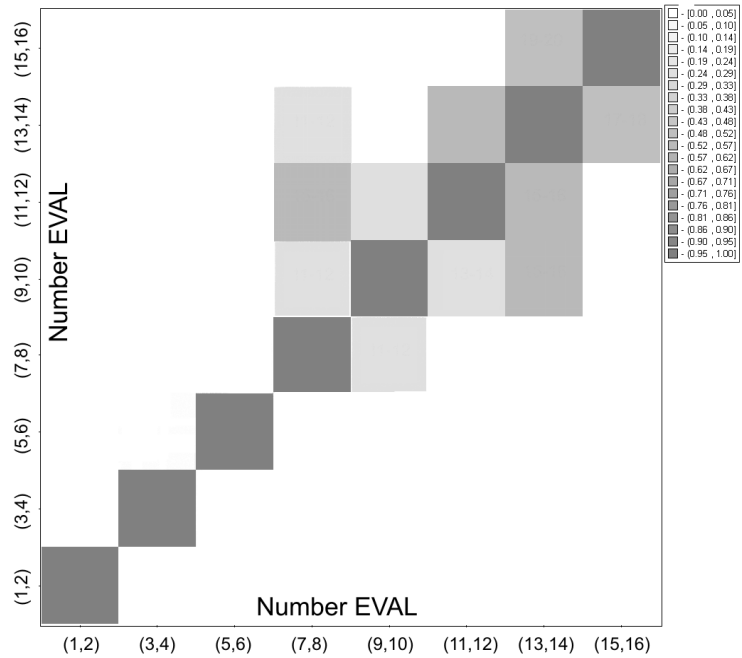

Fig.3. The correlation matrix for the first 16 restored components of the series. Correlations are shown in the 21-st color gray scale from 0 to 1 (scale to the right).

Now, we can restore the form of the sought oscillatory mode. Filtering the trend and noise in accordance with the type of matrix of correlations, we obtain the desired mode (Figure 4a) and its periodogram (Figure 4b ).

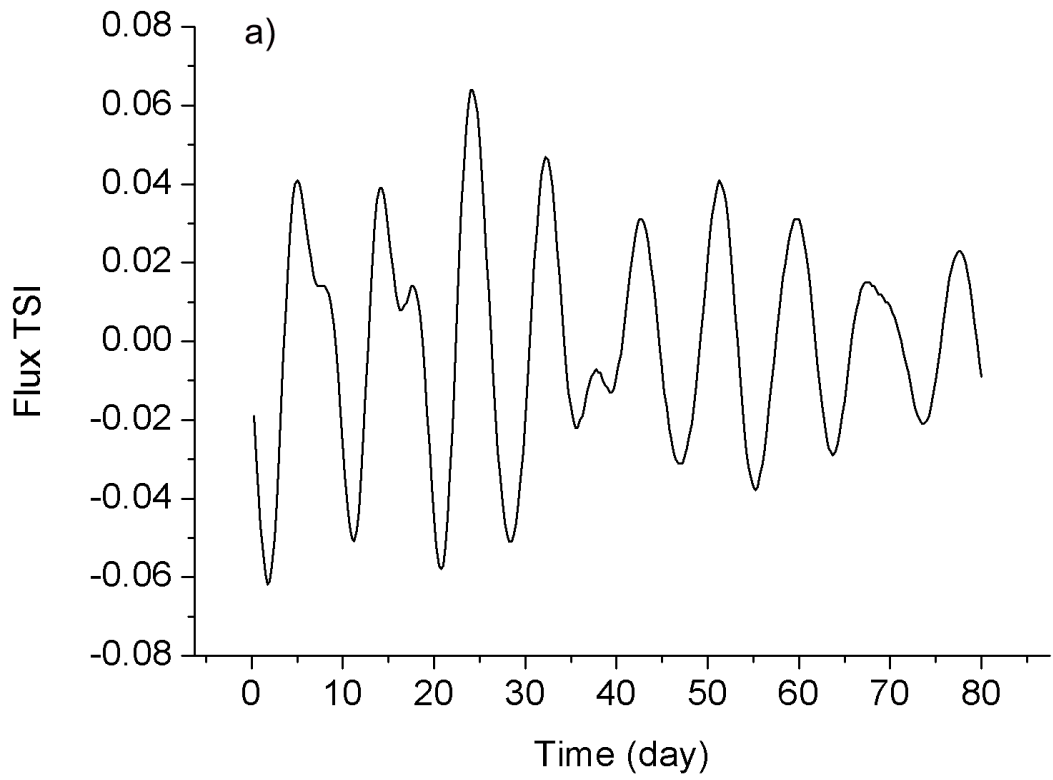

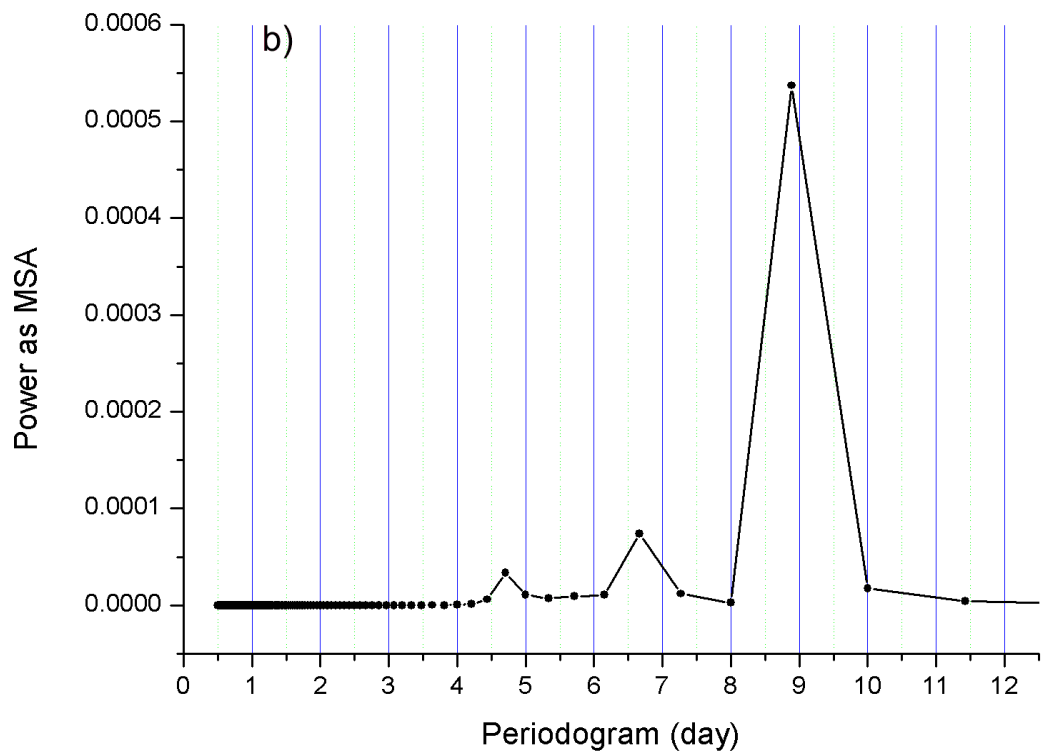

Fig.4. a) The reconstructed oscillatory part of the investigated series, consisting of three independent decomposition components (Number EVAL 1-6); b) its periodogram.

The periodogram shows that the mode consists of three components with periods of 4.7, 6.6, 9 days, among which the 9-day mode is dominant, and the other two modifies it only slightly.

The wavelet transformation (Fig. 5) for the reconstructed oscillatory mode (see Fig. 4 top) does not significantly add information about the spectral composition of the mode, on the contrary: it shows the existence of one dominant mode in 8-10 days and masks the other two. But what is important is that it indicates its stability during the observation period.

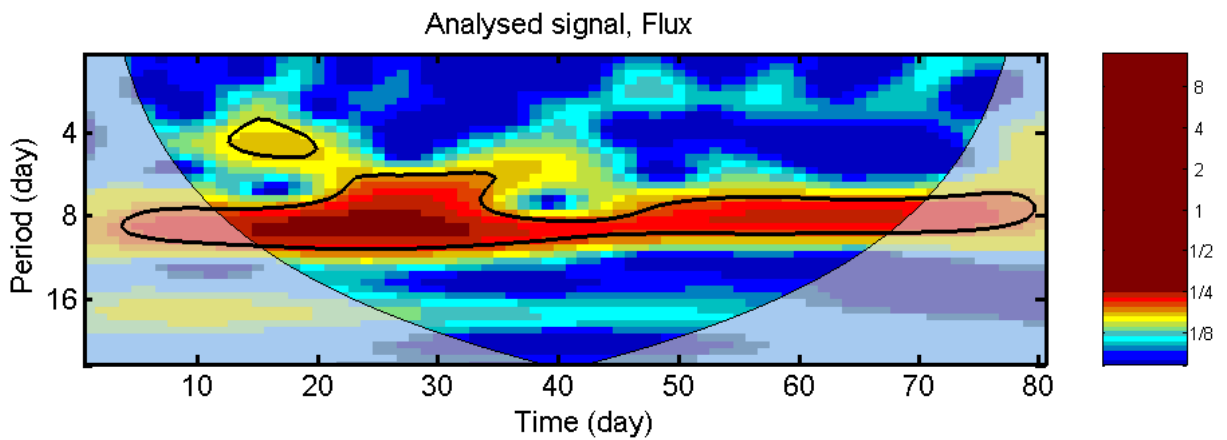

Fig.5. Wavelet transform for the reconstructed oscillatory mode

It is known that in the Total Solar Irradiance (TSI) data, a mode with a period of about 10 days was observed earlier. It is usually assumed that it can be the 3rd harmonic from the rotation frequency of the Sun (Javaraiah, 2011). This explanation seems unconvincing to us, because the mode of 8-10 days is much more stable than the mode of rotation, the period of which varies greatly from 27 to 34 days.

All the oscillation modes obtained by us from the arbitrarily taken fragments of the general series of TSI values are summarized in Table 1. The statistical characteristic, the standard deviation (SD), is calculated for the most stable modes. It can be seen that the 27-33-day mode has a significantly larger spread of SD values than the 9-day mode. This fact indicates that the mode of 9 days is an independent mode, and is not an overtone of the 27-33 day mode.

Tabl. 1

| Obs | Interval, | Modes |
|---|---|---|

|  | day | 36- 38 h | 4-6 d | 9 d | 27-33 d |
|---|---|---|---|---|---|
| 2017_5 apr - 25 jun | 80 | 1.7 | 6.6 | 9 | 27.5 |
| 2014_17 dec-2015_30 sept | 287 | 1.6 | 6.5 | 9.2 | 25 |
| 2007_21 jul-2008_19 dec | 516 | 1.5 | - | 9 | 27 |
| 2008_7 jul - 18 sept | 74 | 1.5 | - | 8.9 | 33.5 |
| Standard Deviation |  | 0.096 | - | 0.12 | 3.66 |

We note that, in general, this lowest-frequency mode is not the subject of our study, since, as we shall see below, the series constructed from reflected planetary sunlight do not have a long duration. Of the outer planets studied here, only for Mars we can construct time series of changes in the solar flux with the duration of the order of the month, due to slow motion this planet in the field of view of the SOHO / LASCO coronagraph.

**3. Investigation of long-period oscillations of the Sun as a star from the average brightness of the photosphere in images obtained with SOHO/MDI and SDO/HMI**

We are looking for weak brightness fluctuations, i.e. temperature, in the integral flux from the Sun. To do this, one can use the SOHO and SDO space data, namely the disk images averaged over the disk in the continuum. This is a satisfactory imitation of the radiation of the Sun as a star. We get a signal of solar brightness oscillations integrated over the disk, which is a superposition of millions of independent luminance modes from each point of the image.

**3.1. SOHO / MDI data**

Let us take a sufficiently long observation, satisfying the static conditions necessary to reveal the significant periodic component of the series. For example, the observation of 1998.02.01_00: 00: 00 - 1998.04.13_00: 00: 00 = 73 days, Cadence = 8 hours.

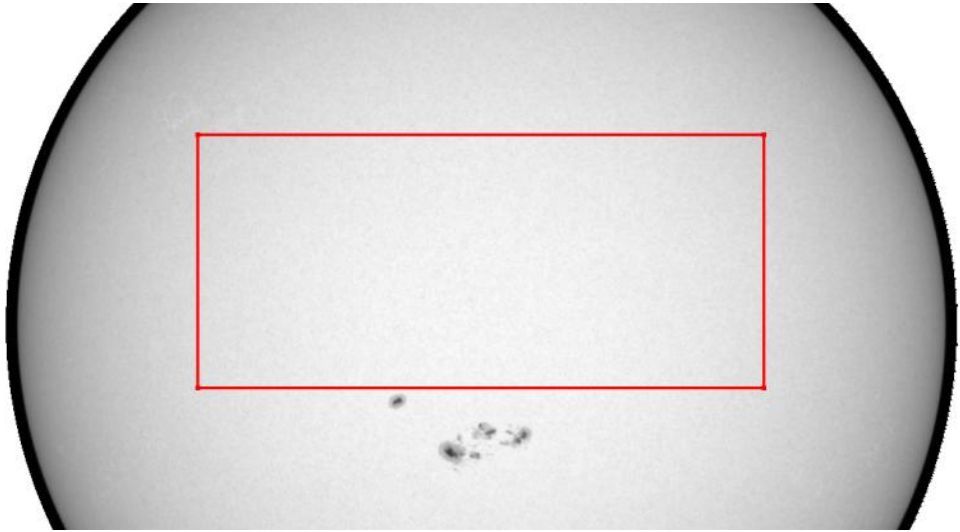

Fig.6. Position of the frame in the size 600x170 ps for SOHO/MDI data.

The frame is selected by such a way that not one spot gets into it during the observation period. Here, of course, we are not guaranteed against accidental entry of flocculent fields that can modify the flux from the area chosen, but, most likely, their influence near the equatorial zone is very weak. As a result, a series of average flux values for the selected area (Fig.7) and phase diagrams of the components of the singular expansion (Fig.8) look as follows:

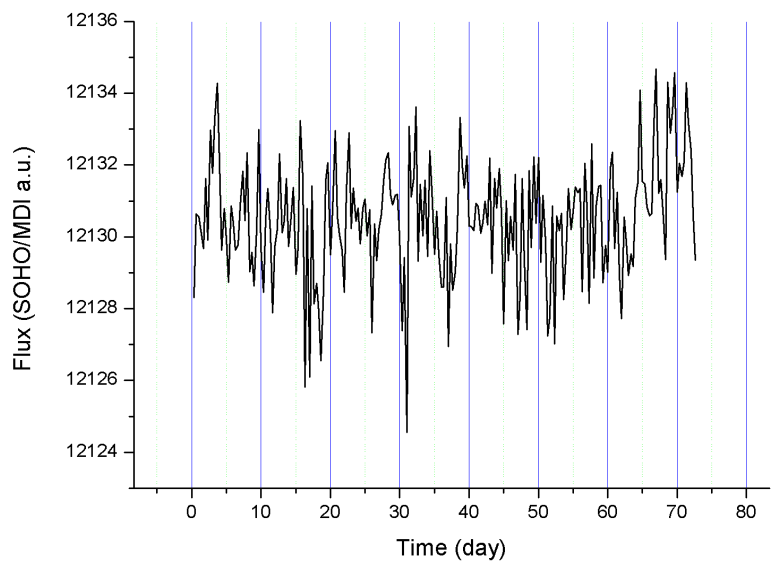

Fig.7. Change in average flux values for the selected area in the period 1998.02.01_00: 00: 00 - 1998.04.15_00: 00: 00. SOHO/MDI.

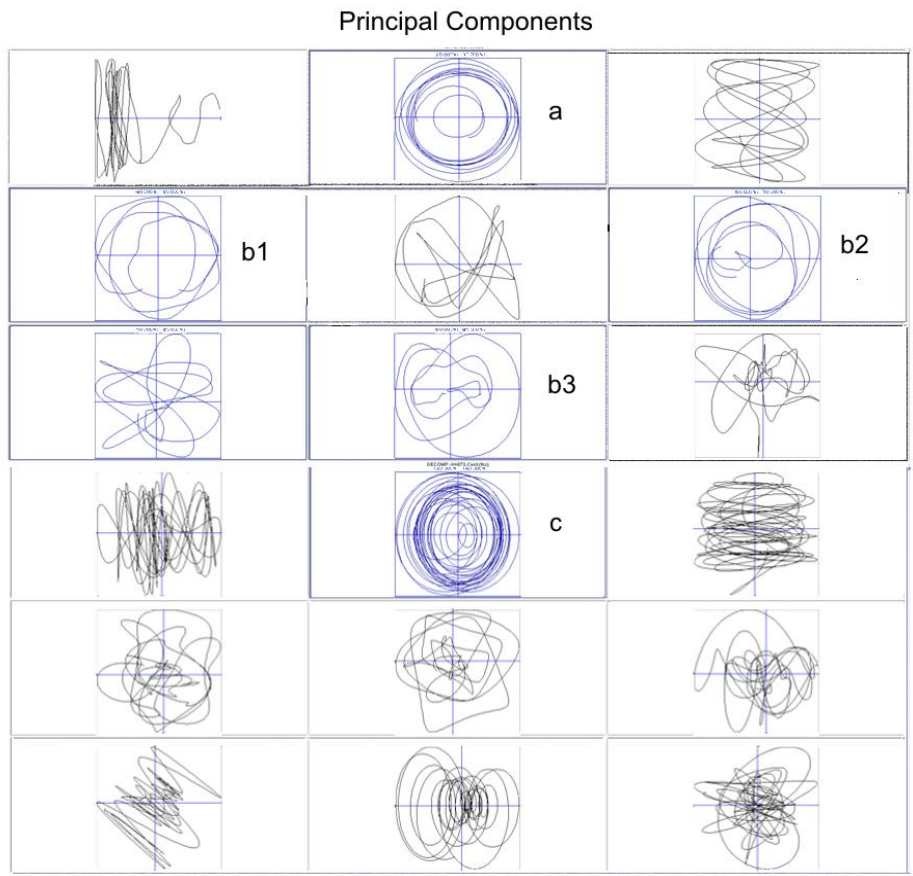

Fig.8. The first 18 phase diagrams for the components of the singular expansion of the investigated series (shown above in Fig. 7), are presented.

Let us pay attention to one feature: the mode (a) with a period of 4.6 days and the mode (c) with a period of 1.6 days (38 hours) have an almost harmonic characteristic, while the 9-day mode, although circulating , but is a bit more complicated (it is represented in the expansion by three components with the same amplitudes). As we noted above, the phase diagrams represent 2-dimensional images of the eigenvectors of the singular expansion of the investigated series and are arranged in decreasing order of amplitudes. In this case, at first glance, we observe some inversion of the amplitudes with respect to the results of point 2.2, but it turned out that the total amplitude of the expansion components of the series for the 9-day mode (b1 + b2 + b3) is not less than the amplitude of the mode "a "(4.6 days), then, as before, there is a mode "c" (with 38 hours). Thus, the contradiction, emerging at first glance, is allowed.

## 3.2. SDO/HMI data

When working with SDO / HMI data, the main difficulty in detection of periodic components is the presence of a daily artifact (T = 24 hours). This artifact has a significant amplitude and masks real cyclic processes, which are contained in time series obtained from SDO / NMI data. However, as we have shown, the stability of this parasitic mode allows it to be filtered out confidently (Smirnova et al., 2013).

For a reliable selection of the desired periodicals (and this is the case of the M36h and M9d modes) from the time series of intensities constructed from SDO / HMI data, we proceed in the same way as above. Let the area under investigation be located in the center of the Sun with a size of 1024x256 ps, along which we will determine the average value of the radiation flux. For the confident allocation of the above modes, the data obtained with a 4-hour cadence are quite appropriate. Let the length of the series be of the same order 70-80 days. Consider the observation - a series of filtergrams obtained in an intensity with the duration of 93 days (558 pts), from April 9 to July 10, 2011. Fig. 9 shows the change in the average flux from the area under study (left) and an arbitrary fragment of this time series (right). Trend, i.e. the general decrease in intensity during the observation time is due to the fact that the Earth (together with the station), in its motion along the orbit, moves away from the Sun. The fragment clearly shows that the artifact - the 24 hour mode is dominant and constitutes the main part of the periodic process. However, it can be seen that its amplitude is slightly modulated (it is not constant, as it should be with a harmonic process), this suggests that there is another perturbing component in the time series under investigation, which we will try to determine.

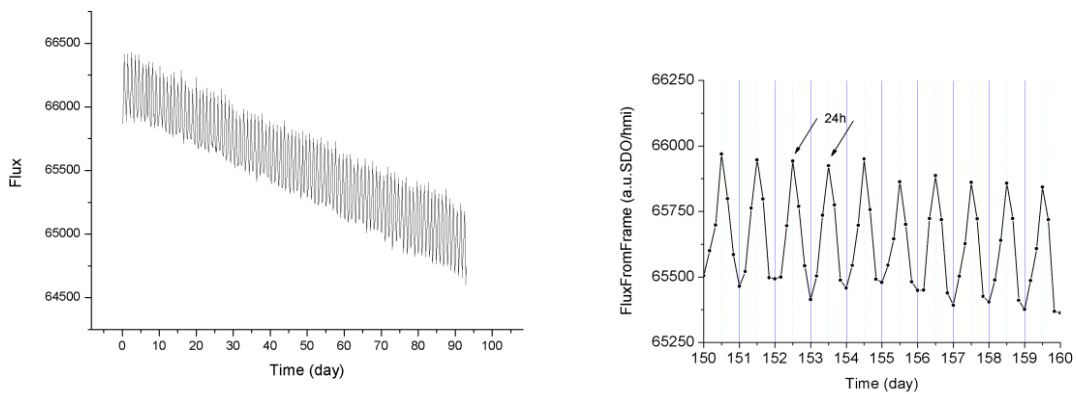

Fig.9. On the left: a change in the average intensity flux on the area during the period from April 9 to July 10, 2011 (93 days). On the right: an arbitrarily selected fragment of the time series.

So far as our sought-for modes (M36h and M9d) have longer periods, we use the simple

Adjacent Average method to filter the parasitic mode, i.e. the window smoothing. SSA - analysis of the residual time series shows that its periodic part consists of 3 independent components: 1,6d, 9d, 25d (Fig.10).

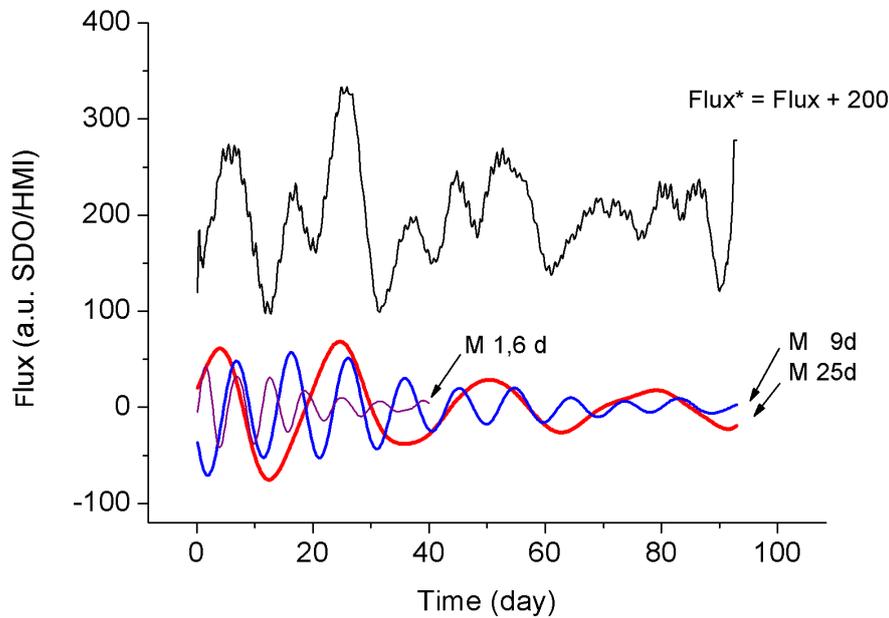

Fig.10. Smoothed and detrended series (the smoothing window is 4 days); with its periodic modes (1.6d, 9d and 25d).

4. **Investigation of long-period oscillations of the brightness of the Sun as a star in the sunlight reflected from the planets during their passage in the field of view of the LASCO C3 coronagraph**

The cosmic coronagraph LASCO C3 (https://lasco-www.nrl.navy.mil/) provides a wonderful opportunity to study global variations in the brightness of the Sun. Gaulme et al.(2016), and Rowe et al. (2017) have used the Kepler K2 space telescope data to monitor the brightness of sunlight reflected from the planet Neptune. In the power spectra, they found periods of approximately 4, 8 and 13 hours, to which they attributed solar origin. This result served as an incentive for our research. To study the long-period variations in the brightness of the Sun as a star, we use the data of the LASCO C3 coronagraph, which, in our opinion, has a number of advantages for the problem being solved before the Kepler K2 telescope observations. The mission "K2" represents a monitoring the bright stars in different parts of the ecliptic for about 75 days. Observation areas are selected by NASA experts to search for exoplanets and, incidentally, Neptune appeared in one of the series. In contrast, in the field of view of the LASCO C3 coronagraph, the planets, especially the inner ones, appear often and predictably. The LASCO C3 is one of the instruments of the SOHO spacecraft and since May 1996 it has obtained almost continuous series of FITS images of the outer solar corona with a cadence of 12 minutes: (ftp://sohoftp.nascom.nasa.gov/qkl/lasco/quicklook/level_05/).

In the field of view of coronagraph, all zodiacal objects that are at the time of observation in the area of the sky with a radius of 3.5 to 30 solar radii centered in the Sun are seen. The SOHO apparatus is located at the Lagrange point L1 of the Earth-Sun system and its data do not have harmful 12-hour and 24-hour artifacts inherent in spacecraft in geosynchronous orbits. The LASCO C3 coronagraph telescope has an input aperture of 110 mm and is equipped with a CCD receiver with 1024 × 1024 pixels. The area of the image is a square with a side of 21.5 mm, each pixel is a square with a side of 0.021 mm. The shutter speed (exposure time) of the LASCO C3 recording chamber is about 19 seconds. In the coronagraph there is vignetting of the inclined rays of light with a lens objective. Since not all oblique rays reach the focal plane, there is a gradual darkening of the image along the edges of the field of view. This manifests itself as a trend in the time series of the planet's brightness, which is easily removed during processing. We did not use those images when the planets were too close to the solar corona and to areas with maximum vignetting and diffraction. It should be noted that LASCO tools are not new. They were built in the late 1980s, when digital cameras were imperfect. Therefore, the images obtained by LASCO C3 coronagraph sometimes exhibit various technical errors: broken lines, pixel dips, etc. Telemetry cutoffs are also observed, caused by radio interference or disturbances in data transmission to the Goddard Space Flight Center. Firmware update was not performed once. For our work, these shortcomings of the source material do not pose any fundamental difficulties. Since we examine the variable part of the signal in the brightness of the sun's light reflected from the planets, possible problems with the calibration and saturation of the signal do not interfere with the detection of quasiperiodic components. The suitability of LASCO data for accurate photometry of stars was demonstrated Sigismondi et al. (2014).

The use of synchronous series of planets and neighboring stars in the field of view of LASCO 3C facilitates the allocation of solar harmonics. There are series, when in the field of view are two planets, for example, Mars and Jupiter. The coincidence of harmonics in power spectra indicates their solar nature. The high brightness of some planets does not prevent the emphasizing of a periodic signal. In the field of view of the LASCO C3 coronagraph near to the investigated planet, it is always possible to select a suitable star for the control. Sometimes a planet can get into a coronal mass ejection, but because of the short duration of the process, this is practically not reflected in the power spectrum. We used the series when no large coronal mass ejections were observed.

The LASCO C3 coronagraph database allows us to perform the comprehensive studies on both inner planets and external ones. As an observational material for detecting periodic changes in the reflected sunlight from the SOHO/LASCO C3 space observatory data archive, FITS sequences of intensities with cadence of 12 minutes and a size of 1024 × 1024 pixels were taken. For the searching for periodicities in the time series we used along with the Fourier analysis, the above described method of investigating non-stationary series and non-linear systems CaterPillarSSA (Golyandina et al.,2001). In this paper, we confined ourselves to the study of outer planets. 5-day continuous time series of brightness of the planets of Mars and Jupiter have been studied. We found two matching modes: 8-10 hours and 36-38 hours, which, apparently, are of solar origin. In addition, 25-day time series of variations of reflected solar radiation was obtained for Mars. In it, an even more low-frequency 10-day mode was discovered. In this case, the amplitudes of the detected modes decrease from low to high frequencies. The found period of time variations of sunlight reflected from Mars practically coincides with the value of the periodicity in TSI. Here we present some examples of processing time series for planets.

**Mars 2015/05/27-31**

Figure 11 shows the detrended time series of solar radiation reflected from Mars in the interval from May 27 to May 31, 2015 (it is expressed in arbitrary flux units of SOHO/LASCO) and three main modes with periods of 8-10, 24-26 and 36-40 hours. The observation interval was 123 hours.

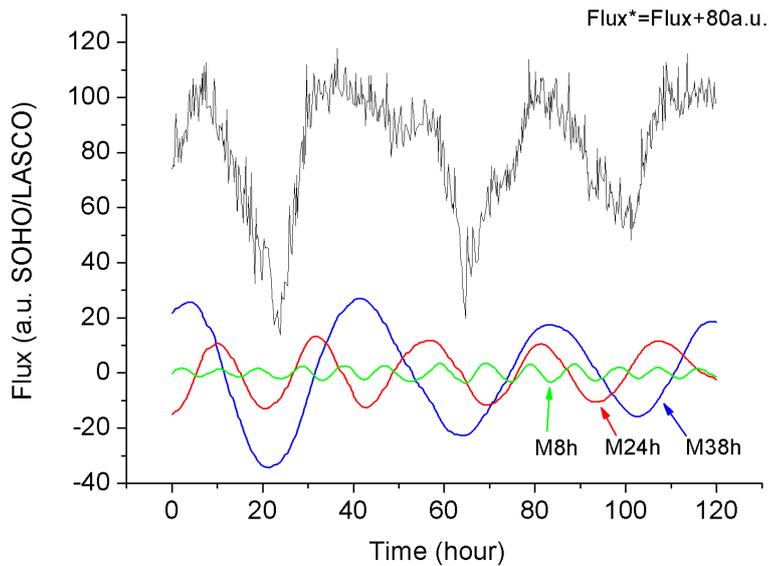

Рис. 11. Detrended time series of variations of the sunlight flux reflected from the planet Mars and its three main decomposition components: modes with periods of 8-10, 24-26 and 36-38 hours.

**Jupiter 2016/09/17-21**

The phase diagrams of the main components of the decomposition of the time series of the variation of the sunlight reflected from Jupiter show that only three quasiperiodic components are present in the original series. Fig. 12 shows the original time series after the removal of the trend and its three main components of decomposition with periods of 36-38 hours, 15 and 8-10 hours, respectively.

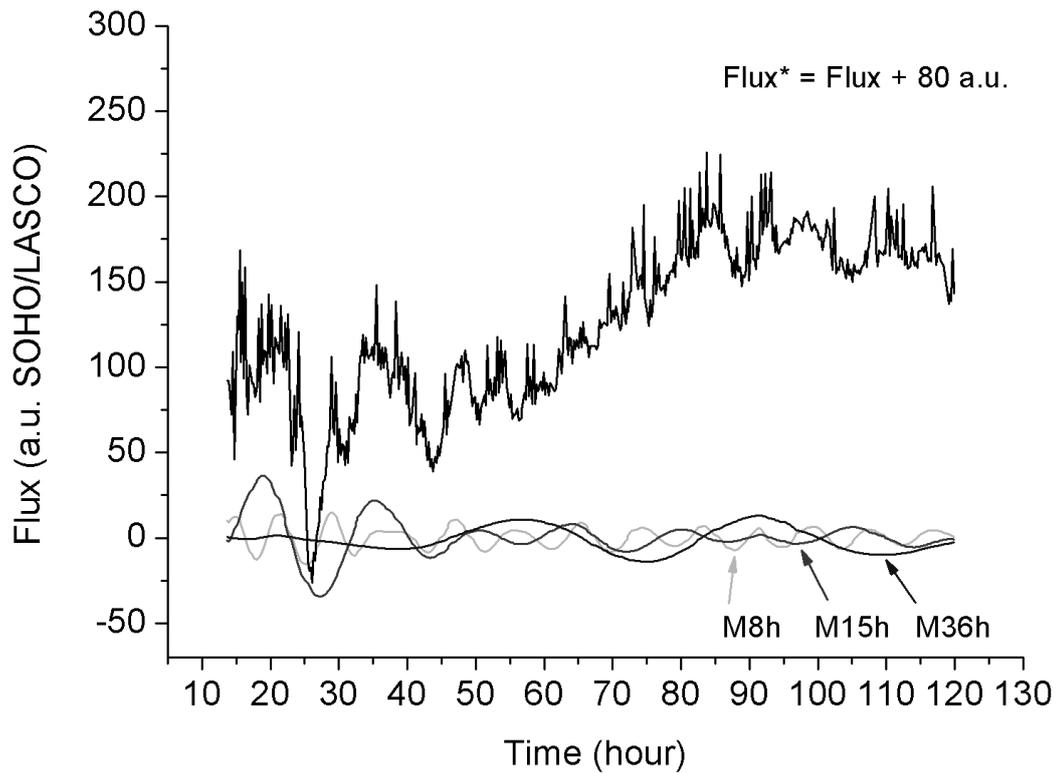

Рис. 12. Detrended time series of the sunlight flux reflected from the planet Jupiter and its three main decomposition components: modes with periods: 8-10, 15 and 36-38 hours.

**Mars  2013/ 30/04-20/05**

In conclusion, we give an example of processing the time series 2013/30 / 04-20 / 05 for Mars with a duration of 527 hours. Mars moves in the field of view of the LASCO C3 coronagraph slower than the stars. This allows us to obtain a sufficiently long series of observations. Figure 13 shows the detrended time series and the three main modes with periods of 8 hours, 36-40 hours and 9-10 days. The amplitudes of the modes found decrease from low to high frequencies.

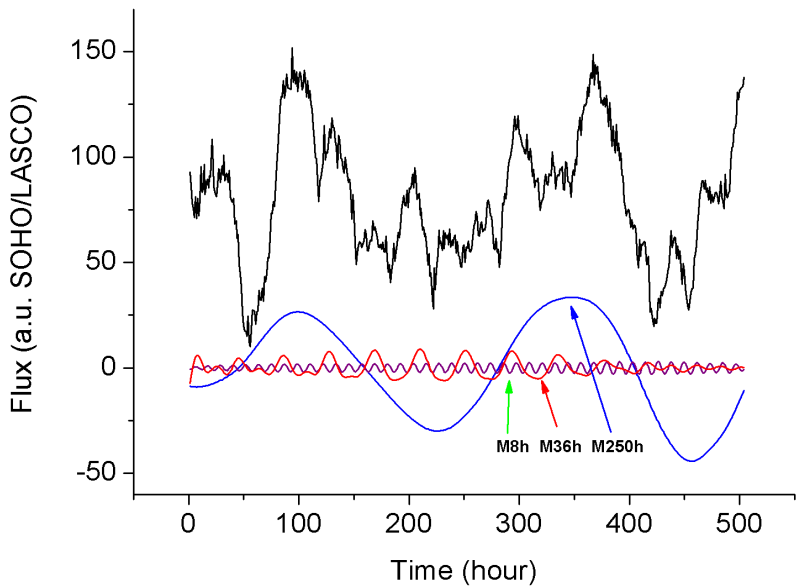

Fig. 13. Detrended time series of the sunlight flux reflected from the planet Mars (upper) and its three main decomposition components: modes with periods: 8, 36-38 and 220-250 hours (8-10 days).

To check the obtained results for the presence of an artifact, the synchronous series for control stars and a nearby small area of background were formed. Figure 14 shows the synchronous time series for the Pollux star (the lower curve) and the average values for the flux on the background area of 15 × 20 pixels, located below Pollux - the upper curve (it is shifted upward along the Y axis for the convenience of consideration). It is clearly seen from the figure that in the background area, there are no oscillations of brightness completely.

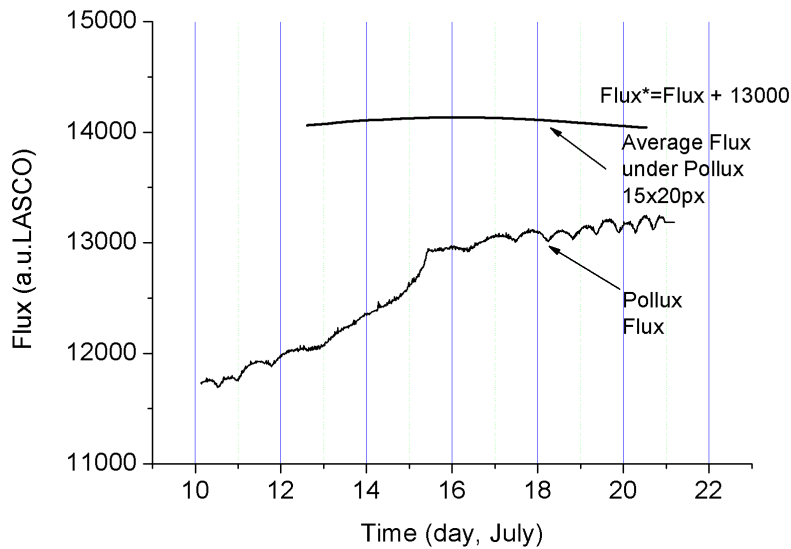

Fig. 14. Synchronous time series for the Pollux (lower curve) and mean values for the background area 15x20 px taken near the Pollux - upper curve.

Thus, in 5-day continuous time series of the brightness of the planets of Mars and Jupiter, two coincident modes were found: 8-10 and 36-38 hours, which, apparently, are of solar origin. For Mars, the 25-day time series of reflected solar radiation was obtained, in which a 10-day mode has been detected.

5. **Conclusions**

We present the results of processing various observational data for studying low-frequency modes (with periods from tens of minutes to tens of days) in global variations of the brightness of the Sun as a star, using the time series of TSI, averaged images of the photosphere and time variations of sunlight reflected from planets. Significant periods of 8-10 hours, 36-40 hours and 9-10 days have been identified. Perhaps, the period of 9 days is associated with the rotation of the splitted kernels of lowest gravitational modes, found by Fossat et al. (2017).

The authors are grateful to the SOHO and SDO teams for the opportunity to use the observational data.
 The work is supported by the Program № 28 of the Presidium of RAS and the RFBR grant 18-02-00168. A.S. thanks Russian Scientific Foundation for the support (project 15-12-20001.

**References:**
Astafyeva, N. M.; Zastenker, G. N.; Eiges, P. E., Wavelet Analysis of Fluctuations in the Flux of Solar Wind Ions, 1996, Cosmic Research, Vol. 34, No. 4, p.375

Broomhead D.S., King G.P. Extracting qualitative dynamics from experimental data // Physica D: Nonlinear Phenomena, Volume 20, Issue 2-3, p. 217-236. (1986). Vol. 20. C. 217-236. DOI: 10.1016/0167-2789(86)90031-X

Fossat et al., Asymptotic g modes: Evidence for a rapid rotation of the solar core, 2017, Astronomy & Astrophysics, 604, A40, DOI: 10.1051/0004-6361/201730460

Foullon et al. , Ultra-long-period Oscillations in EUV Filaments Near to Eruption: Two-wavelength Correlation and Seismology, 2009, The Astrophysical Journal, Vol. 700, Issue 2, pp. 1658-1665), DOI: 10.1088/0004-637X/700/2/1658

Gabriel, A. H., Grec, G., Charra, J., et al., Global Oscillations at Low Frequency from the SOHO Mission (GOLF), 1995, Sol. Phys., 162, 61, DOI: 10.1007/BF00733427

Gaulme P., et al., A Distant Mirror: Solar Oscillations Observed on Neptune by the Kepler K2 Mission, 2016, http://arxiv.org/abs/1612.04287v1

Golyandina, N., Nekrutkin, V., and Zhigljavsky, A., 2001, Analysis of Time Series Structure: SSA and Related Techniques, London: Chapman & Hall/CRC

Kennedy, J. R., Jefferies, S. M., & Hill, F., 1993, Solar G-Mode Signatures in P-Mode Signals, in GONG 1992, Seismic Investigation of the Sun and Stars, ed. T. M. Brown (San Francisco: ASP), ASP Conf. Ser., 42, 273

Kopp, Greg; Lean, Judith L., A new, lower value of total solar irradiance: Evidence and climate significance ,2011, Geophysical Research Letters, Volume 38, Issue 1, CiteID L01706, DOI: 10.1029/2010GL045777

Rowe J. et al., Time-Series Analysis of Broadband Photometry of Neptune from K2, 2017, http://arxiv.org/abs/1702.02943v1

Severnyi A.B., Kotov V.A., Tsap T.T., 1976, Observations of solar pulsations. Nature 259: 87–89.

Sigismondi C., et al., Photometry of Delta Scorpii from 1996 to 2013 using SOHO LASCO C3 coronograph, 2014, https://arxiv.org/abs/1410.8492

Smirnova, V.; Efremov, V. I.; Parfinenko, L. D.; Riehokainen, A.; Solov'ev, A. A., Artifacts of SDO/HMI data and long-period oscillations of sunspots, 2013, Astronomy & Astrophysics, Volume 554, id.A121, 7 pp., DOI: 10.1051/0004-6361/201220825